\documentclass[acmsmall,nonacm]{acmart}

\AtBeginDocument{%
  }

\setcopyright{acmlicensed}
\copyrightyear{2025}
\acmConference[CHI PLAY 2025 Workshop]{Exploring Future Directions for Healthy Escapism and Self-Regulation in Games}{October 13, 2025}{Pittsburgh, USA}
\acmBooktitle{CHI PLAY 2025 Workshop: Exploring Future Directions for Healthy Escapism and Self-Regulation in Games}
\acmYear{2025}
\acmDOI{10.1145/XXXXXXX.XXXXXXX}

\begin{document}

\title[If You Hold Me Without Hurting Me]{If You Hold Me Without Hurting Me: Pathways to Designing Game Audio for Healthy Escapism and Player Well-being}
\titlenote{Presented and discussed at the CHI PLAY 2025 Workshop \textit{Exploring Future Directions for Healthy Escapism and Self-Regulation in Games}, Pittsburgh, USA, October 13, 2025.}

\author{Caio Nunes}
\email{caioeduardo@alu.ufc.br}
\orcid{0000-0001-5997-3518}
\affiliation{%
  \institution{Federal University of Ceará}
  \city{Fortaleza}
  \state{Ceará}
  \country{Brazil}
}

\author{Bosco Borges}
\email{boscoborges4@gmail.com}
\orcid{0000-0002-8644-1055}
\affiliation{%
  \institution{Federal University of Ceará}
  \city{Fortaleza}
  \state{Ceará}
  \country{Brazil}
}

\author{Georgia Cruz}
\email{georgia@virtual.ufc.br}
\orcid{0000-0002-9127-0529}
\affiliation{%
  \institution{Federal University of Ceará}
  \city{Fortaleza}
  \state{Ceará}
  \country{Brazil}
}

\author{Ticianne Darin}
\email{ticianne@virtual.ufc.br}
\orcid{0000-0003-3617-5462}
\affiliation{%
  \institution{Federal University of Ceará}
  \city{Fortaleza}
  \state{Ceará}
  \country{Brazil}
}
\renewcommand{\shortauthors}{Nunes et al.}

\begin{abstract}
Escapism in games can support recovery or lead to harmful avoidance. Self-regulation, understood as combining autonomy with positive outcomes, is key to this distinction. We argue that audio, often overlooked, plays a central role in regulation. It can modulate arousal, mark transitions, and provide closure, yet its contribution to well-being remains underexplored. This paper identifies methodological and accessibility gaps that limit recognition of audio’s potential and outlines ways to address them. We aim to encourage researchers and developers to integrate audio more deliberately into the design and study of healthier escapist play. 
\end{abstract}



\received{20 February 2025}
\received[revised]{12 March 2025}
\received[accepted]{5 June 2025}

\maketitle

\section{Introduction}

Escapism has long been recognized as one of the main motivations for gameplay, appearing across models of player motivation and media use \cite{yee2006motivations, klimmt2009video, caplan2009problematic}. It is often described as the tendency to detach from everyday concerns or stressors, yet its consequences are ambivalent. On the one hand, escapism can offer distraction, stress relief, and opportunities for affective regulation \cite{bowditch2018coping}. On the other hand, it can reinforce avoidance and interfere with everyday functioning. Recent work emphasizes this tension by distinguishing between healthy escapism, when play supports recovery and well-being, and subversive escapism, when it undermines them \cite{kosa2020four}. This perspective shifts the debate from whether escapism is good or bad to a more pressing question: \textit{"Which factors determine when escapist play sustains well-being and when it becomes harmful?"}

One of the most decisive factors in this distinction is self-regulation. In Self-Determination Theory (SDT), self-regulation refers to the extent to which behavior is experienced as autonomous, reflecting choice and volition rather than external pressure or compulsion \cite{ryan2017self}. Applied to games, this perspective captures whether players can autonomously decide when to begin, how long to continue, and when to disengage. Research consistently shows that difficulties in regulation are linked to excessive or compulsive gaming \cite{kardefelt2014conceptual}. At the same time, successful self-regulation allows games to function as bounded recesses that facilitate recovery, coping, and positive mood management \cite{reinecke2009games, kosa2020four}. Framing escapism through this lens highlights \textit{autonomy} as a key condition for distinguishing between healthy and subversive play. However, a player could autonomously choose to play for a duration that negatively impacts their life responsibilities. In this case, the behavior is self-determined, but the outcome is detrimental. Hence, a healthy escapism also requires \textit{positive outcomes}, where psychological relief (e.g., reported mood improvement) is supported by physiological recovery (e.g., stress marker reduction), independent of the feeling of volition during play. Consequently, we draw attention to how the design of game elements can either scaffold a player's ability to self-regulate or undermine it.

\textbf{We argue that audio is a key game element that can be intentionally designed to structure the escapist experience itself, facilitating autonomous self-regulation and positive outcomes}. While often perceived as secondary to visual or mechanical elements, audio possesses unique qualities that make it a potent regulatory channel. Unlike visual cues, which require focused attention, sound is pervasive and processed with a lower cognitive load, capable of influencing arousal and emotional states subconsciously, creating a direct tension between systemic regulation and player autonomy that design must address. The well-documented link between music and physiological responses, from emotion regulation and stress recovery to everyday relaxation \cite{kamioka2014effectiveness, koelsch2015music}. This makes audio a prime channel for influencing the very recovery and emotion regulation that are central to healthy escapism \cite{kosa2020four}. Besides, audio's ability to structure time and rhythm transitions offers a subtle yet material means of scaffolding player awareness without disrupting immersion.

Yet, despite its unique potential, the role of audio in player well-being remains poorly understood. Much of the game-audio literature concentrates on immersion, presence, arousal, and affect, seldom linking these outcomes to well-being or self-regulatory processes \cite{nunes2024echoes}. This gap is especially evident in discussions of escapism and self-regulation, which often emphasize constructs such as absorption, affective modulation, stress relief, and distraction \cite{calleja2010digital, stenseng2021there}. While these emphases are not identical, they share a concern with how players become deeply involved and how their emotional states are shaped during play.

This lack of synthesis leaves a critical gap, preventing designers and researchers from systematically using audio as a tool for player well-being. This position paper moves to fill that gap. We contend that audio is not a peripheral feature but a central, designable mechanism for scaffolding healthy escapism in games. To substantiate this claim, the remainder of this paper will first identify methodological and accessibility gaps previously identified in game audio research that perpetuate this problem. We then propose three concrete directions for future work focused on using audio to support self-regulation and positive emotional outcomes.

\section{How Research Gaps Obscure Audio's Role in Healthy Escapism}

In a previous study, we reviewed how audio and player experience are evaluated \cite{nunes2024echoes}. The results show a fragmented evidence base where game audio is routinely linked to immersion and affect, but its deeper impact is unclear. A key problem is that self-report measures often fail to converge with physiological data; for example, players may report feeling calm while their arousal markers remain elevated \cite{nunes2024echoes}. These inconsistencies make it challenging to draw conclusions about outcomes such as coping, stress recovery, or emotion regulation. This methodological fragmentation is compounded by other persistent issues, including a lack of personalization and the tendency for audio to be overshadowed in high-sensory contexts \cite{rogers2019audio, rogers2018vanishing, prechtl2014methodological, ribeiro2020game}. Together, these findings reveal that audio is seen as an immersive aesthetic but not perceived as a potential lever for self-regulatory outcomes.

The real-world impact of this research gap is starkly visible in games designed for sensitive audiences. Our literature review (n=49) on games for autistic players \cite{borges2025sounds}, for instance, shows that audio is often absent from game design descriptions and from user evaluation, despite its potential to support engagement or, if mishandled, trigger overload in audiences with auditory hypersensitivity. Across the surveyed studies, only 14\% (n=7) explicitly evaluated audio; when they did, its uses ranged from background music to movement sonification, yet the impacts on the target audience were seldom reported in depth \cite{ragone2020osmosis}. Such an absence of systematic evaluation and reporting reveals a double gap: a lack of design transparency and potentially a neglect of accessibility. For that audience, without customization options or sensitivity to diverse auditory needs, audio risks becoming a barrier instead of a resource. These findings highlight how sound design, if overlooked, can miss therapeutic opportunities and expose players to overstimulation or discomfort. This case starkly illustrates how neglecting audio can undermine a player's ability to self-regulate, turning a potentially healthy escape into a subversive, dysregulating experience.

While these examples are not exhaustive, the gaps identified in the literature prevent us from measuring whether audio supports well-being or causes distress. Methodological blind spots mean that we lack ways to evaluate if audio is contributing to the positive outcome of healthy escapism, whether audio supports coping and recovery or simply intensifies immersion. Accessibility gaps can imply that sound becomes overwhelming or exclusionary, rather than restorative. In both cases, escapism risks sliding toward avoidance and overuse -- not because audio cannot support healthier outcomes, but because research and design fall short in providing the tools to differentiate them. The absence of variation, closure, and inclusive controls leaves players to self-manage through workarounds such as muting or replacing game soundtracks \cite{rogers2019audio}, undermining the potential of audio to contribute directly to constructive escapist experiences.

Ultimately, players are left without the necessary scaffolds to exercise their autonomy due to a lack of key tools, such as methods for measuring player responses to audio cues or designs that utilize audio to create natural stopping points. For instance, if safe exit points are signaled by a shift to calmer music in a hub area, it creates a clear closing motif after a mission, or even a brief \textit{"breathing window""} of silence after a moment of high tension, all of which provide a sense of closure that could help players disengage with a sense of accomplishment. Without such intentional design, poorly tested or repetitive audio can instead reinforce compulsive immersion or force avoidance, both of which undermine self-regulation. These blind spots leave players without the necessary scaffolds to disengage autonomously, showing how overlooked audio can shift escapism from recovery to overuse.

\section{Towards Self-Regulation and Healthy Escapism with Audio}

The gaps highlighted reveal a significant methodological deficiency: despite audio's clear influence, we remain unable to reliably determine its impact on recovery versus its potential to undermine regulation. Strategies are absent that repurpose sound as a design utility for player empowerment and healthy boundaries. To resolve this issue, we propose three core directions focusing on addressing methodological pitfalls, enhancing reporting transparency, and solidifying collaborative ties between research and industry.

First, adopt multi-modal evaluation: self-reports, behavioral traces, and psychophysiological markers each provide partial insights, but their combination offers a fuller picture of audio’s regulatory role. Logs of muting, playlist swaps, or sudden volume drops can reveal when players regulate their own auditory load. Coupled with physiological markers and short probes, such as \textit{“Do you feel calmer?”} or \textit{“Are you ready to stop?}” one can test whether sound genuinely supports coping or merely shifts physiology without providing felt relief. These behaviors can act as practical, scalable proxies for laboratory measures. Convergences and divergences are equally informative, allowing researchers to identify when design succeeds and giving developers concrete indicators of overload or recovery.

Second, report design intentions transparently. Findings remain anecdotal when studies only state that sound was present. Reporting must clarify how audio was crafted and with what regulatory purpose to subconsciously guide players toward recovery, or to provide explicit cues for conscious self-regulation: a low-tempo loop added after combat to support recovery, a brief silence window following repeated failures to encourage pause, or a closing motif marking the transition to a safe hub area. By grounding technical decisions in regulatory intent, researchers can accumulate evidence across contexts and developers can align design choices with player well-being, especially in sensitive cases where unexamined audio risks overstimulation \cite{borges2025sounds}.

Third, foster a shared agenda between research and industry. Rather than prescribing isolated techniques, advancing this field requires collaboration and teamwork. Initiatives such as joint workshops, repositories of customizable regulatory audio patterns and adaptive sound systems, or co-produced guidelines, similar to existing accessibility standards, can foster cumulative knowledge.

Together, these commitments elevate audio from a mere backdrop to an active mediator of escapist play. They reposition sound as a critical lever for discerning the balance between recovery and over-engagement, underscoring that without systematic evidence, designers still lack the means to use it intentionally for player well-being. To drive this agenda, we invite the community to consider:

\begin{enumerate}
    \item Which combinations of self-report, behavioral, and physiological measures best capture audio’s regulatory role?
    \item How might soundscapes, finality cues in UI sounds, and music be designed to make transitions between immersion and disengagement more intentional?
    \item What forms of player control over audio foster genuine autonomy in shaping escapist experiences?
    \item Which standardized or recurring sound patterns are being used unreflectively, and how might their effects on well-being be better understood?
\end{enumerate}

\section*{Author Biographies}

\noindent\textbf{Caio Nunes} is a researcher with a Master’s degree in Computer Science from Federal University of Ceará, Brazil. His work focuses on Human–Computer Interaction (HCI) and Game User Research, exploring audio, musical visualizations, motivation, and well-being in emotional communication.

\vspace{0.6em}
\noindent\textbf{Bosco Borges} is a Master’s student at Federal University of Ceará, Brazil. His work focuses on HCI and Game User Research, examining how sound and interaction design can support player experience, promote healthier engagement, and cater to diverse audiences.

\vspace{0.6em}
\noindent\textbf{Georgia Cruz} is an Adjunct Professor at Federal University of Ceará, Brazil. Her work focuses on UX, Audiovisual, and Communicability, exploring user experience, gender in games, and critical play, with an emphasis on how interactive media can foster reflection and influence well-being.

\vspace{0.6em}
\noindent\textbf{Ticianne Darin} is an Associate Professor of HCI at the Federal University of Ceará. Her game research examines the foundational components of player experience, from motivation and engagement to the pursuit of eudaimonia through reflexive games. She is currently focused on developing ethical alternatives to manipulative design to foster digital well-being. Her leadership in the Brazilian research community includes chairing the nation's primary HCI symposia and founding the WIPlay workshop series, a dedicated venue to unite the country's HCI and games researchers. As a UNESCO consultant, she helps shape national policy on digital rights, focusing on implementing "Child and Adolescent Rights by Design" frameworks for games and online services.

\bibliographystyle{ACM-Reference-Format}
\bibliography{sample-base}

\end{document}